\documentclass{aadebug}

\usepackage{xspace}
\usepackage{psfrag}
\usepackage{paralist}
\usepackage{enumerate}
\usepackage[sort]{cite}
\usepackage{listings}
\newcommand{\sref}[1]{Sec.~\ref{#1}}
\newcommand{\fref}[1]{Fig.~\ref{#1}}

\newcommand{\eg}{e.g.\xspace}

\newcommand{\Layer}[2]{\textsf{Layer$\,{#1}_{#2}$}\xspace}
\newcommand{\Layers}[2]{\textsf{Layers$\,{#1}_{#2}$}\xspace}
\newcommand{\Lzeros}{\Layer{0}{s}}
\newcommand{\Lzerom}{\Layer{0}{m}}
\newcommand{\Lzerosm}{\Layers{0}{s/m}}
\newcommand{\Lone}{\Layer{1}{m}}
\newcommand{\Ltwo}{\Layer{2}{m}}
\newcommand{\Lthree}{\Layer{3}{m}}
\newcommand{\Lzeroserver}{L$0_{m}\,$Server}
\newcommand{\Loneserver}{L$1_{m}\,$Server}

\newcommand{\tool}[1]{\textsf{#1}\xspace}

\newcommand{\Fiddle}{\tool{Fiddle}}
\newcommand{\FGI}{\tool{FGI}}
\newcommand{\PADI}{\tool{PADI}}

\newcommand{\TV}{\tool{TotalView}}
\newcommand{\PPDD}{\tool{p2d2}}

\newcommand{\GDB}{\tool{GDB}}
\newcommand{\DBX}{\tool{DBX}}
\newcommand{\DDBG}{\tool{DDBG}}

\newcommand{\Deipa}{\tool{Deipa}}
\newcommand{\TeSS}{\tool{TeSS}}
\newcommand{\STEPS}{\tool{STEPS}}

\newcommand{\PVM}{\tool{PVM}}

\newcommand{\XML}{\tool{XML}}

\newcommand{\echoclient}{\texttt{echo\_client}\xspace}
\newcommand{\echoserver}{\texttt{echo\_server}\xspace}

\newcommand{\mainthread}{\emph{main thread}\xspace}
\newcommand{\managerthread}{\emph{manager thread}\xspace}

\newcommand{\launcherthread}{\emph{launcher thread}\xspace}

\newcommand{\pvmspawn}{\texttt{pvm\_spawn()}\xspace}
\newcommand{\Pvmspawn}{\texttt{Pvm\_spawn()}\xspace}
\newcommand{\Pvmexit}{\texttt{Pvm\_exit()}\xspace}

\begin{document}
                                                     
\corr{0309027}{143}

\runningheads{João Lourenço, José C. Cunha and Vitor Moreira}{Control and Debugging of Distributed Programs Using Fiddle}

\title{Control and Debugging of Distributed Programs Using Fiddle}

\author{
João Lourenço\addressnum{1}\comma\extranum{1},
José C. Cunha\addressnum{1},\\
Vitor Moreira\addressnum{1}
}

\address{1}{
  Departamento de Informática \\
  Faculdade de Ciências e Tecnologia \\
  Universidade Nova de Lisboa \\
  Portugal
}

\extra{1}{E-mail: \{jml, jcc, vrm\}@di.fct.unl.pt}

\pdfinfo{
/Title (Control and Debugging of Distributed Programs Using Fiddle)
/Author (João Lourenço, José C. Cunha and Vitor Moreira)
}

\begin{abstract}
  The main goal of \Fiddle, a distributed debugging engine, is to
  provide a flexible platform for developing debugging tools. \Fiddle
  provides a layered set of interfaces with a minimal set of debugging
  functionalities, for the inspection and control of distributed and
  multi-threaded applications.
  
  This paper illustrates how \Fiddle is used to support integrated
  testing and debugging. The approach described is based on a tool,
  called \Deipa, that interprets sequences of commands read from an
  input file, generated by an independent testing tool. \Deipa acts as
  a \Fiddle client, in order to enforce specific execution paths in a
  distributed \PVM program.  Other \Fiddle clients may be used along
  with \Deipa for the fine debugging at process level.  \Fiddle and
  \Deipa functionalities and architectures are described, and a
  working example shows a step-by-step application of these tools.
\end{abstract}


\section{Introduction}
\label{sec:introduction}

Developing parallel and distributed applications is still a difficult
task, even after about several decades of intense research on
methodologies and support tools.
  On one hand, this is due to the intrinsic complexity of distributed
computing, involving many dynamic interacting entities, executing on
multiple processors, leading to a global nondeterministic behavior. On
the other hand, much research and development in this area has been
facing a need to continuously adapt to new hardware, operating system
and middleware platforms, and new programming language approaches
(ranging from imperative to functional or logical, and to object- and
agent-oriented models). Such constant evolution in the computing
models and platforms made many interesting tools obsolete, as they
were not able to adapt, or their design was too dependent upon
specific models or platforms.

As parallel machines were becoming available since the 80s, parallel
debuggers have been developed and some of them evolved into
significant commercial tools, such as \TV~\cite{TotalView}.
Initially, such debuggers usually followed a monolithic approach,
where a debugging engine and its user interface were combined into a
single large program.

Developments on distributed computing have motivated debugging
architectures based on the client/server model, such as
\PPDD~\cite{p2d296}, which clearly separates the user interface from
the debugging engine.  Most designs are based on having a separate
process in each machine node, to support the remote access to the
debugging functionalities, and rely upon a central server to do all
the processing. This was an important design decision to enable the
debugging tool to adapt to different user requirements and
environments but the importance of such a step was not recognized
until recently. Related and significant efforts were also done
concerning monitoring of parallel and distributed
programs~\cite{hpcn:clvmv98,euro98,omis,ocm,first,hpcn2000,atempt}.

An important requirement for flexible development tools is their
ability to adapt to new computing platforms or user requirements. This
motivates an approach based on a clear separation between the software
layers which support the minimal functionalities required by a tool,
and the layers which provide the extensions that may be required by
specific user and application scenarios. This requires a debugging
support platform that enables the inclusion of new tools, to provide
complementary functionalities for application development and acting
as intermediaries between the debugger and the other tools.

In previous work, some of the above requirements, involving the
integration of a testing tool for parallel programs (\STEPS) and a
parallel debugger (\DDBG) were tested~\cite{euromicro:lckknw97}.
However, the limitations of such experiment have motivated the
continuation of this work towards a more flexible debugging
architecture, \Fiddle.

In this paper, we first describe the main characteristics of the
\Fiddle architecture. Next, in \sref{sec:testing-and-debugging}, we
discuss an approach for integrated testing and debugging and, in
\sref{sec:example} we illustrate the approach through a working
example.  Then, in \sref{sec:deipa}, we describe how \Deipa was
implemented as a \Fiddle client tool, and we present conclusions and
future work in \sref{sec:conclusions}.

\section{The Fiddle Distributed Debugging Engine}
\label{sec:fiddle}

\Fiddle~\cite{fiddle,iccs:2001} is a distributed debugging engine
based on a client-server operational model.  The debugging engine acts
as a server and the debugging tools, which are not included in \Fiddle
specification, act as the client(s).

The \Fiddle debugging engine has the following main characteristics:

\begin{itemize}
\item \emph{Support for interactive correctness debugging.}  \Fiddle
  includes all the traditional debugging facilities available in
  sequential debuggers, such as \DBX~\cite{dbx} and \GDB~\cite{gdb},
  but extended to operate upon the multiple processes of a distributed
  program.  Some of the services provided by \Fiddle operate at
  process level, \eg, breakpointing and single-stepping, while some
  other operate at application level, \eg, obtaining control of newly
  created application processes (\emph{spawn'ed} in a \PVM program).
  As a whole, \Fiddle provides a consistent and complete set of
  services which would be unavailable if a simple set of independent
  sequential debuggers would have been used instead;
  
\item \emph{Client/server model.}  Multiple debugging interfaces
  (client tools) can connect simultaneously to \Fiddle and have access
  to the same set of target processes, providing, in this way,
  multiple ``debugging views'' over the same target application;
  
\item \emph{Extensibility.}  \Fiddle functionality may easily be
  extended by adding new clients, which provide specific and/or
  complementary functionalities to the debugging system;
  
\item \emph{Support of high-level user-defined abstractions.}  Besides
  the ability to debug distributed programs at textual (source) level,
  \Fiddle can also be applied to programs developed using higher
  level programing languages and/or models, such as visual parallel
  programming languages, which allow a parallel application to be
  specified in terms of a set of graphical entities describing
  application components and their interconnections.  \Fiddle
  extensions can be incorporated into the debugging system, which maps
  the functionalities required by such high-level debugging interface
  onto \Fiddle basic debugging services;
  
\item \emph{Tool synchronization.} By allowing multiple concurrent
  client tools to access a common set of target processes, \Fiddle
  needs to provide some basic support for tool synchronization. Using
  \Fiddle this can be achieved in two ways:
  \begin{inparaenum}[i)]
  \item as all requests made by client tools are controlled by a
    central daemon, \Fiddle is able to serialize the service requests,
    avoiding some of the interferences among these tools;
  \item client tools can receive notification events reporting changes
    in the target processes and/or service requests issued by other
    clients, allowing a tool to be aware of the ``debugging
    environment'', and to react upon changes\footnote{This
      coordination mechanism is not implemented in \Fiddle yet.  See
      \sref{ssec:fiddle-current-status}.}.
  \end{inparaenum}
  
\item \emph{Easy integration in Parallel Software Engineering
    Environments.}  \Fiddle extensibility can be explored to support
  the cooperation and integration of debugging and other related tools
  in the environment, \eg, on-line program visualization tools.
\end{itemize}

\subsection{\Fiddle Software Architecture}
\label{ssec:fiddle-soft-arch}

\begin{figure}[htbp]
  \centering
  \psfragscanon
  \psfrag{CT13m}{{\small CT$_2^{3_m}$}}
  \psfrag{CT23m}{{\small CT$_1^{3_m}$}}
  \psfrag{CT12m}{{\small CT$_1^{2_m}$}}
  \psfrag{TP1}{{\small TP$_1$}}
  \psfrag{TP2}{{\small TP$_2$}}
  \psfrag{TP3}{{\small TP$_3$}}
  \psfrag{Layer 0s}{{\small \Lzeros}}
  \psfrag{Layer 0m}{{\small \Lzerom}}
  \psfrag{Layer 1m}{{\small \Lone}}
  \psfrag{Layer 2m}{{\small \Ltwo}}
  \psfrag{Layer 3m}{{\small \Lthree}}
  
  \includegraphics[width=0.7\textwidth]{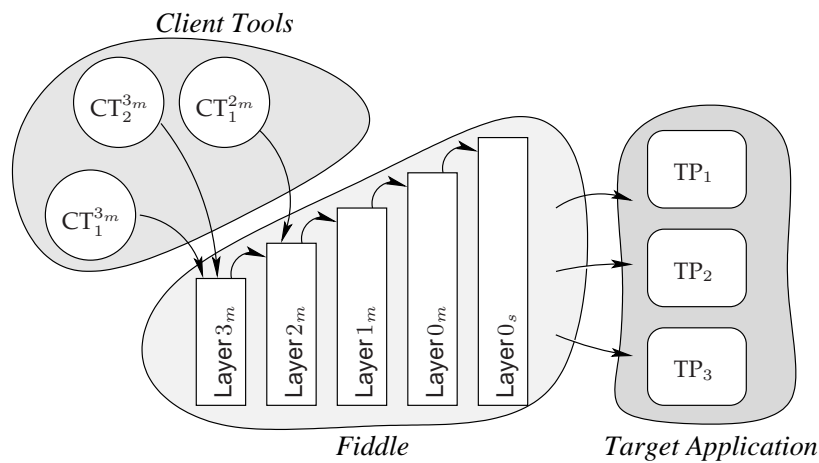}
  \caption{\Fiddle's Layered Architecture}
  \label{fig:layered-architecture}
\end{figure}

\Fiddle is structured as a hierarchy of five functional layers, each
providing a specific set of debugging services and accessible through
an interface library.  Any layer may be used directly by a client tool
if its set of services is found to be adequate for the client needs.
Each layer is also used indirectly by the layer immediately above.
For example, in \fref{fig:layered-architecture}, \Ltwo has two
clients: the tool CT$_1^{2_m}$ and \Lthree.

A service requested by, \eg, a \Lthree client tool, will be processed
and transferred successively to the underlying layers, until \Lzeros
is reached, and the service requested applied to the target process.
The reply to such service request is also successively processed and
transferred to the upper layers until the client tool gets the result.

There is a minimum set of functionalities common to all \Fiddle
layers, namely:

\begin{itemize}
\item \emph{Inspect/control multi-threaded target processes.}  \Fiddle
  provides a basic set of debugging services to act upon threads
  within a multi-threaded process;
  
\item \emph{Inspect/control multiple target processes concurrently.}
  Any client tool may use \Fiddle services to inspect and control
  multiple processes concurrently;
  
\item \emph{Support for client tool(s).}  All layers accept, at least,
  requests from one client tool or from the layer above. The upper
  layers also accept multiple client tools operating concurrently upon
  the same target application.
\end{itemize}

\begin{figure}[htbp]
  \centering
  \psfragscanon
  \psfrag{FIDDLE 0s}{\footnotesize \Lzeros}
  \psfrag{FIDDLE 0m}{\footnotesize \Lzerom}
  \psfrag{FIDDLE 1m}{\footnotesize \Lone}
  \psfrag{FIDDLE 2m}{\footnotesize \Ltwo}
  \psfrag{F0s-Lib}{\tiny \Lzeros Library}
  \psfrag{F0m-Lib}{\tiny \Lzerom Library}
  \psfrag{FIDDLE-1m Library}{\tiny \Lone Library}
  \psfrag{FIDDLE-2m Library}{\tiny \Ltwo Library}
  \psfrag{F0m Server}{\scriptsize \Lzeroserver}
  \psfrag{F1m Server}{\scriptsize \Loneserver}
  
  \includegraphics[width=0.8\textwidth]{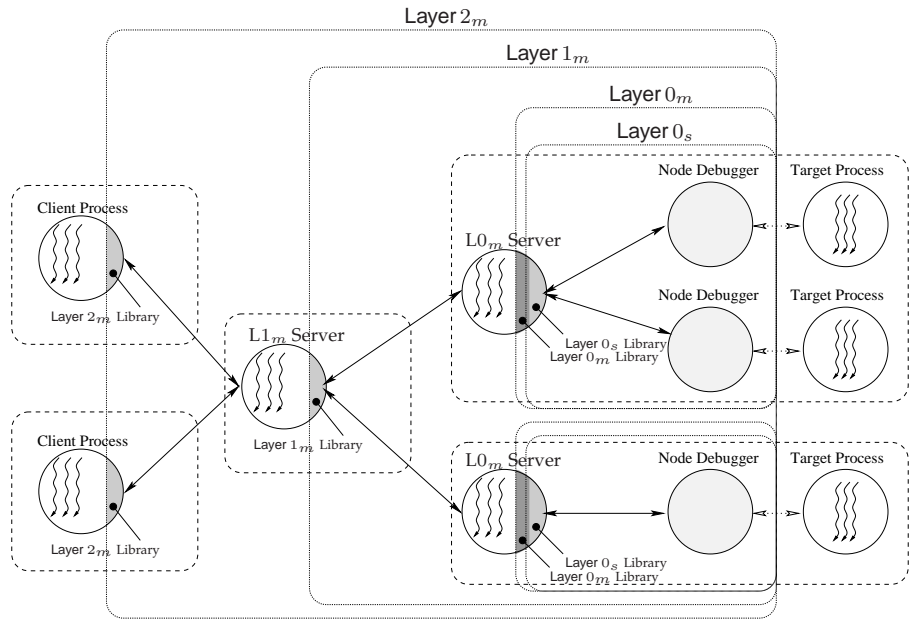}
  \caption{The \Fiddle software architecture}
  \label{fig:all-layers}
\end{figure}

Besides these common functionalities, each layer provides a set of
specific functionalities.  These functionalities are supported by the
software architecture shown in \fref{fig:all-layers}.

\Fiddle layers numbers range from \emph{zero} to \emph{three}, with
incremental functionality, and have a suffix \emph{s} or \emph{m},
representing the class of accepted clients: \emph{single-} or
\emph{multi-threaded}, respectively.

\paragraph{\Lzeros}
This layer implements a set of local debugging services, including the
common set of debugging services described above, which are made
available to a single client tool.

The following components are known to this layer:
\begin{inparaenum}[i)]
\item \emph{Target processes}: a subset of processes that compose the
  target application and are being executed in the local node;
\item \emph{Node debuggers}: associated to each target process there
  is a node debugger, \eg, a \GDB instance, which is responsible for
  acting upon it;
\item \emph{\Lzeros library}: provides single-threaded access to the
  debugging functionalities;
\item \emph{Client tool}: a single-threaded tool accessing \Fiddle and
  possibly implementing a user interface.
\end{inparaenum}

\paragraph{\Lzerom}
This layer extends \Lzeros to provide support for a single
multi-threaded client tool.  The client may now use threads to
concurrently control multiple target processes and the interaction with
the user.

\paragraph{\Lone}
Both \Lzerosm have access to processes on the local node.  \Lone
extends \Lzerosm to allow a client tool to have access to remote
target processes.

The components known to this layer are:
\begin{inparaenum}[i)]
\item \emph{Target processes}: running in the local or remote nodes;
\item \emph{Node debuggers}: one associated to each target process;
\item \emph{\Lzeroserver}: an instance of this daemon will be
  automatically launched in each node hosting target processes;
\item \emph{\Lone Library}: provides thread-safe access to local and
  remote debugging functionalities;
\item \emph{Client tool}: client tools using this layer may
  transparently access local and remote target processes by using
  their global identifiers.
\end{inparaenum}

\paragraph{\Ltwo}
This layer extends \Lone to provide support for multiple simultaneous
client tools, which may issue concurrent requests to \Fiddle.  This is
achieved by using a \emph{\Loneserver} daemon, as an intermediary
between the client tools and the \emph{\Lzeroserver} daemons.  This
\emph{\Loneserver} daemon multiplexes the service requests from the
multiple client tools, submits them to \Lone, and demultiplexes the
corresponding replies back to the clients.

\paragraph{\Lthree}
\Layers{0/1/2}{} services use a blocking semantics, keeping the
calling thread waiting for the reply. This layer extends \Ltwo to
provide an event-based interaction between \Fiddle and its clients.
In contrast to the lower layers, a thread that invokes a method in the
\emph{\Lthree Library} doesn't block waiting for its reply.  Instead,
it receives a \emph{Request Identifier} which will be used later to
request and process the reply.  When the service is executed, its
success status and data (the \emph{reply}) is sent to the client as an
event.  These events may be processed by the client tool in two
different ways:
\begin{inparaenum}[i)]
\item \emph{Asynchronous mode}: the general methodology to process
  \Fiddle events is to define a handler.  When a service is executed
  by \Fiddle, a new thread will be created to execute the handler.  A
  structure describing the results of the requested service is passed
  as an argument to the handling function, together with the
  \emph{Request Identifier};
\item \emph{Synchronous mode}; in this mode, \Fiddle keeps the
  notification of the event pending until the client tool explicitly
  requests it by invoking a \Fiddle primitive.
\end{inparaenum}

\subsection{\Fiddle Current Status}
\label{ssec:fiddle-current-status}

Currently, \Fiddle \Layers{0/1/2}{s/m} are fully implemented and
functional.  Implementation of \Lthree, which will provide event
notification and event handling mechanisms for client tools, is
ongoing work.  \Fiddle development has taken part on Linux machines,
but there are no strong dependencies on this operating system.  Also,
as the \Fiddle debugging engine has no user interface, there are no
dependencies on graphical environments/packages either.

If the target application is using \PVM to spawn (launch) new
processes, \Fiddle can also automatically acquire such newly created
processes, launching them under the control of a node debugger, and
making them possible targets for future services requests.

Along with the \Fiddle debugging engine, a set of text oriented
debugging interfaces (\Fiddle consoles) are being developed, one for
each \Fiddle layer, which allow to explore the full set of services
provided by each layer.  Additionally, other \Fiddle client tools have
been developed, such as \FGI~\cite{fgi} (the \Fiddle Graphical
Interface) based in the Gnome Desktop Environment, \PADI~\cite{padi}
(a group oriented parallel debugger) based in the Java AWT and \Deipa
(described in this paper).

Being still in a prototype stage, \Fiddle is currently not publicly
available (yet).  However, it is available on a demand basis to anyone
who request it to the author\footnote{E-mail:
  \texttt{jml@di.fct.unl.pt}}.

\section{Integrating Testing and Debugging: the DEIPA Approach}
\label{sec:testing-and-debugging}

Program correctness requires that a given specification of the
intended behavior be satisfied.  Although the development of
high-level abstractions for distributed programming has contributed to
ease the task of program development, there are still many
opportunities for programming errors, posing the need for support
tools.

Many programming errors can be detected, in a more or less automatic
way, through a static analysis of the program source text. Such
analysis can also assist the programmer in the prediction of the
program behavior concerning specific correctness properties. However,
the program source text is not always available. Also, the global
program behavior is often the result of a combination of behaviors,
depending on the operating system or runtime support layers of a
computing environment. 

This explains the significance of approaches for dynamic analysis,
which are centered upon the observation of real execution. Such
approaches assume, from the beginning, the incomplete nature of this
process, due to the usually huge number of computation states which
can be generated by distributed program execution. As such, they are
typically based on the selection of a finite set of representative
test configurations, followed by an observation of the results of
program execution.  The definition of such test scenarios is, of
course, dependent upon the classes of errors or program properties
that one is trying to check.

The main goal of a debugging tool is to help analysing erroneous
program behavior, possibly identified by a previous testing stage, and
to assist the programmer in the formulation or confirmation of
hypotheses on the causes of errors, and in the tracing of their
origins in the program text.

Testing and debugging are naturally intertwined. On one hand, testing
helps identifying errors whose causes must later be traced in a
debugging stage. On the other hand, after a successful debugging
session, one typically needs to reconsider the set of test
configurations. Debugging can also help identifying unforeseen
situations which may require the design of new testing scenarios.

The above aspects have been recognized for a long time, leading to
many proposals of methodologies and tools for combined testing and
debugging. See \cite{book:krawczyk99,book} for a more complete survey
on this topic.

\begin{figure}[htbp]
  \centering
  \includegraphics[width=1\textwidth]{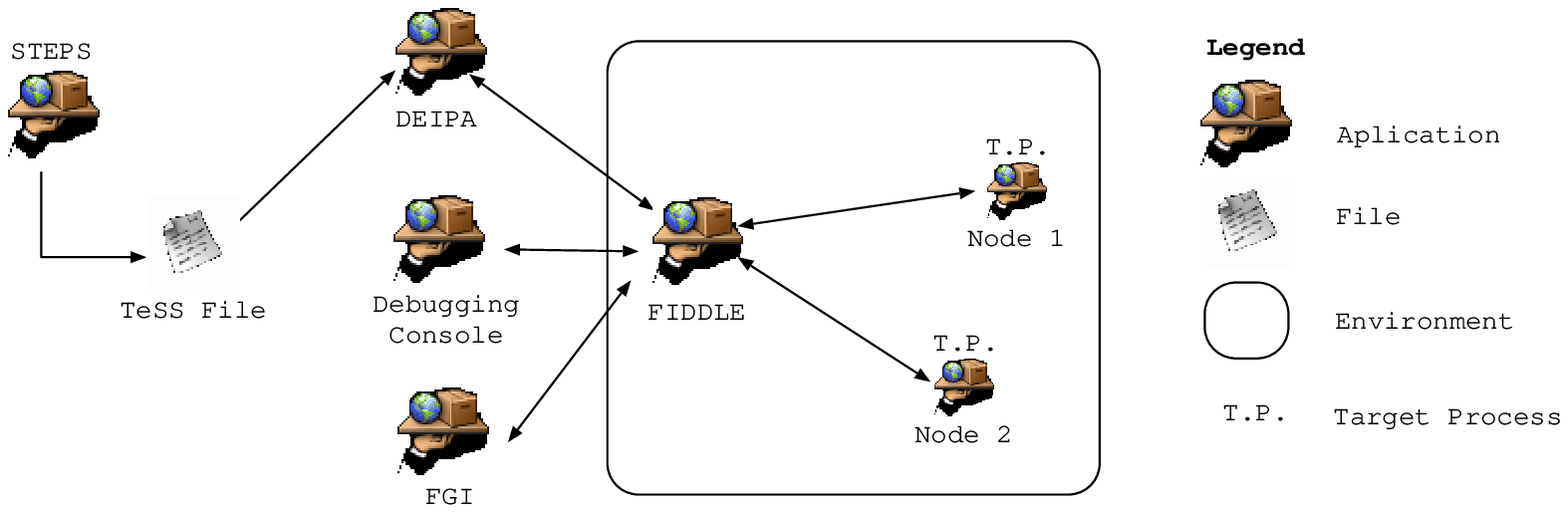}

  \caption{Tool composition of \STEPS and \Fiddle using \Deipa}
  \label{fig:deipa-environment}
\end{figure}

\Deipa~\cite{euromicro:lckknw97} was initially developed to support  a 
testing-and-debugging development cycle, by allowing the composition
of two separately developed tools,
\STEPS~\cite{steps:kw96ii} and \DDBG~\cite{jsa:1999,cai:1998}. The current 
configuration of such testing and debugging architecture is
illustrated in \fref{fig:deipa-environment}.

Based on a static and dynamic analysis of the target program, \STEPS
generates a behavior specification file, the \TeSS file.  This file
includes the definition of a list of \emph{global breakpoints} (a
consistent collection of local breakpoints).  Some of those
breakpoints have special instructions to modify one or more variable
values. This is necessary to drive correctly the application execution
when in presence of conditional branches or loops, \eg, on a
\texttt{if} statement.

To force the application execution to conform to the specification in
the \TeSS file, \Deipa generates sequences of debugging (control and
inspection) commands to \Fiddle, setting a local breakpoint in each
target process and individually driving them until the breakpoints are
reached.  Once a process is stopped in a breakpoint, its internal
status (e.g., variable contents) may be changed, according to the
\TeSS specification.

At any time, the \Fiddle ability to handle multiple clients may be
explored, and the user may switch to another tool, such as \Fiddle's
debugging console or graphical interface (\FGI~\cite{fgi}) and perform
a more detailed inspection and control of the target processes. This
allows the user to perform debugging at distinct abstraction levels,
e.g., at a global program level to inspect process interactions, and
at local process level, to inspect individual process execution.

On completing such a fine (process-level) debugging, the user may
return to \Deipa and proceed with the controlled execution mode, or
release/stop the target application if no more debugging is needed.

\section{A Working Example}
\label{sec:example}

To demonstrate how \Deipa works, we have built a simple application.
It is composed of two programs: the \echoclient, which sends a number
to the server and waits for its reply; and the \echoserver, which gets
a number from the client and exits if the received number is even, or
sends -1 back otherwise.

Since it is common for distributed applications to have a program that
initializes the environment, we consider that the \echoclient is the
startup program, launching the \echoserver server via \pvmspawn.

First, we shall present the source code of both \PVM programs, with
\echoclient on the left column and \echoserver on the right one.  The
tiny leftmost number indicates the line number, used for future
reference.

\lstset{
  language=C
}
\begin{center}
  \begin{tabular}{cc}
  \textbf{\echoclient} & \textbf{\echoserver} \\
  \begin{minipage}{0.55\linewidth}
\begin{lstlisting}
#include <pvm3.h>
#include <stdio.h>
#include <stdlib.h>
#include <unistd.h>



int main ()
{
  int mytid;
  int totid=0;
  int value;


  mytid = pvm_mytid ();

  value = pvm_spawn ("echo_server", NULL,
            PvmTaskDebug, ".", 1, &totid);

  /* Sending 0 will force the server to
     exit, and the client will wait
     forever for the reply */

  value = 0;
  pvm_initsend (0);
  pvm_pkint (&mytid,1,1);
  pvm_pkint (&value,1,1);
  pvm_send (totid, 1);

  /* Get the reply from server */
  pvm_recv (-1,-1);
  pvm_upkint (&value,1,1);
  printf ("Received value %d\n", value);

  pvm_exit();

  return (0);
}
\end{lstlisting}
  \end{minipage}
  &
  \begin{minipage}{0.20\linewidth}\small
\begin{lstlisting}
#include <pvm3.h>
#include <stdio.h>
#include <stdlib.h>


int main ()
{
  int mytid;
  int from, value;
  
  mytid = pvm_mytid();
  
  pvm_recv (-1, -1);
  pvm_upkint (&from,1,1);
  pvm_upkint (&value,1,1);
  
  if ((value%2)==0)
    exit(0);
  else
    value=-1;

  pvm_initsend (0);
  pvm_pkint (&value, 1, 1);
  pvm_send (from, 1);
 
  pvm_exit();  

  return 0;
}
\end{lstlisting}
  \end{minipage}
\end{tabular}
\end{center}

The most relevant source lines in the client code are line~17, where
the server is launched (spawned); line~28, where the message is sent to
the server; and line~31, where the client blocks waiting for the
reply.

On the server, the most relevant source lines are line~13, where the
server waits for a message from the client; line~17, where the server
decides how to react to the received message (terminate or reply); and
line~24, where the server replies to the client (if this was the case).

As the development of the \STEPS tools was discontinued by its
authors, and \Deipa uses \TeSS files as one of its inputs, we produce
the following \TeSS files ourselves, which describe a testing scenario
for the echo client and server processes.  The tiny line numbers are
not part of the \TeSS file itself but were included only for future
reference to specific lines.

\lstset{
  morekeywords={START_FILE,SPAWN_TABLE,INITIAL}
}
\begin{center}
    \hspace{2em}
\begin{lstlisting}
START_FILE:
    echo_client

SPAWN_TABLE:
    {
        0   0   0   0   1   echo_client echo_client.c   13142 4,
        1   16  0   1   2   echo_server echo_server.c   59634 2
    }

INITIAL:
    [{ (1,1,17) }],
    [{ (1,1,28) }],
    [{ (2,1,28) }],
    [{ (2,1,28),(1,2,13) }],
    [{ (2,1,28),(2,2,13) }],
    [{ (2,1,28),(1,2,17,[2,2,"value","1"]) }],
    [{ (2,1,28),(2,2,17) }],
    [{ (1,1,31),(1,2,24) }],
    [{ (2,1,31),(2,2,24) }],
    [{ (1,1,35),(1,2,26) }],
    [{ (2,1,35),(2,2,26) }];
\end{lstlisting}
\end{center}

Each line represents a global breakpoint, as a collection of local
breakpoints of the form ``\texttt{(T,I,L)}'', which are the type of
breakpoint, the virtual identifier of the process (VID) and the line
number respectively.  The type of breakpoint indicates if the program
should stop before (code~1) or after (code~2) the selected line
number. The virtual identifier of the target process will be mapped by
\Deipa to \PVM \emph{task IDs} at runtime.  The new process, at
line~14 of the \TeSS file, is captured by the \emph{launcher} program
and integrated into the \Fiddle and \Deipa execution environment, as
described in \sref{sec:deipa}.

The \TeSS file specifies global breakpoints in the following points:

\begin{itemize}
\item \Pvmspawn, to ensure that new processes are, indeed, created on
  the \PVM environment;
\item Send and receive messages, to enable the verification of
  emission and reception of messages;
\item \Pvmexit, to ensure that all processes get out correctly.
\end{itemize}

\subsection{Using Deipa and Fiddle to capture errors}

The main idea of this example is to launch the distributed application
and control its execution from the \Deipa console; and to examine
individual processes state on the \Fiddle text user interface
(\texttt{f2m} console).  We must notice that this example requires the
usage of \Ltwo, as there are multiple client tools simultaneously
connected to \Fiddle, namely \Deipa, the \emph{launcher} and a \Fiddle
console.

The screen shots in Figures~\ref{fig:capture},~\ref{fig:forcing}
and~\ref{fig:receive} refer respectively to the following situations:
\begin{itemize}
\item The capture of the \echoserver target process, created at
  line~17 of \echoclient, and its integration into \Deipa and \Fiddle;
\item Driving the application behavior by forcing its execution path; and
\item The reception of the answer at the client side.
\end{itemize}

\subsubsection{The capture of the new process}

This task consists of intercepting the spawn action, the launching of
the new target processes under the control of a node debugger, and the
notification of \Deipa and \Fiddle on the existence of the new target
process.

\begin{figure}[htbp]
  \centering
  \includegraphics[width=1.0\textwidth]{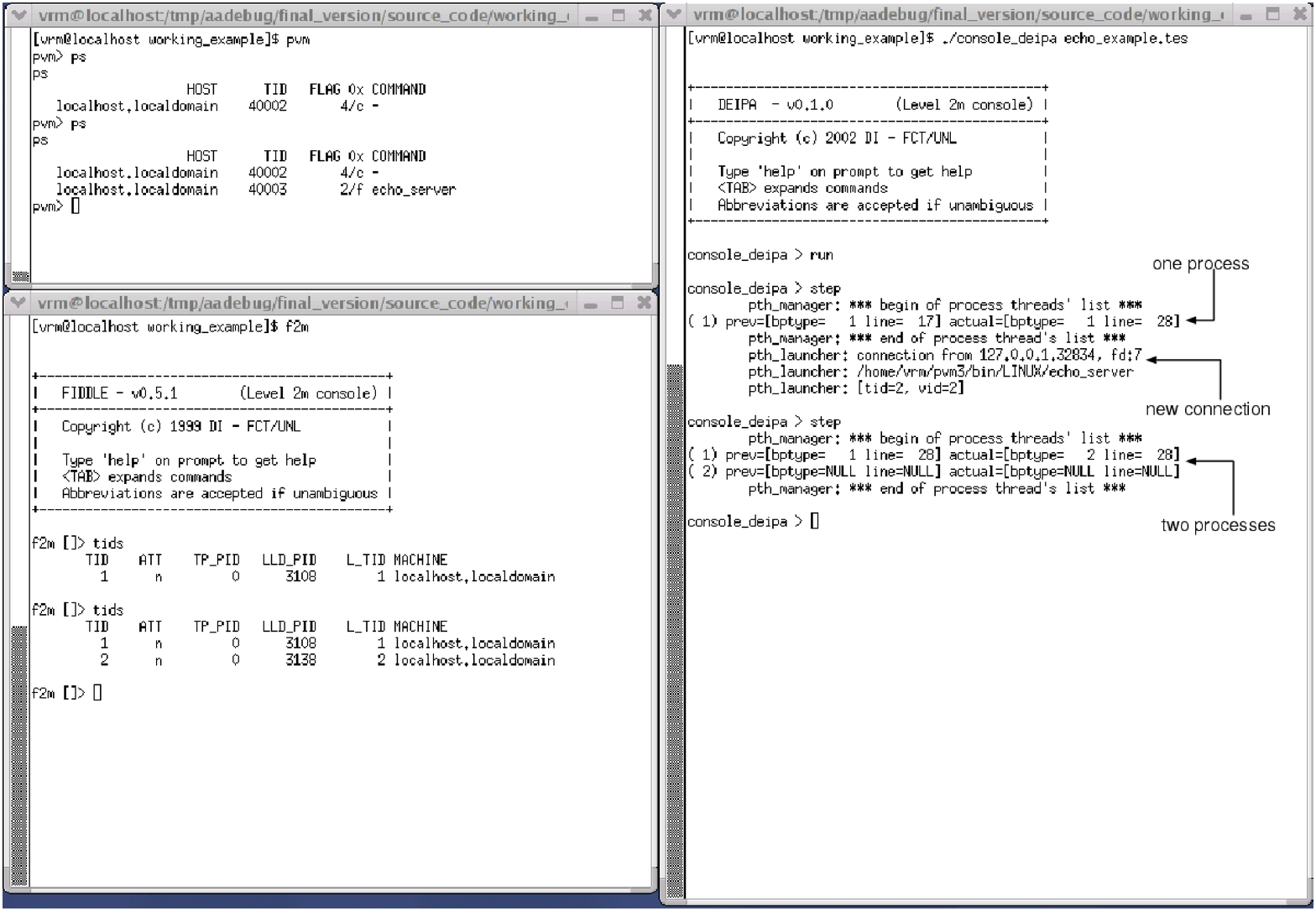}

  \caption{The capture of a new process}
  \label{fig:capture}
\end{figure}

Figure~\ref{fig:capture} shows three windows, the \PVM console on the
top left, the \Fiddle \Ltwo console below, and \Deipa on the right.
On each window two states are shown.  On the first state, the
\echoserver process hasn't been launched yet; while on the second,
when the new process was already started (although not running).

Two of the commands available on the \Deipa console are: \texttt{run},
to start execution the distributed application, and \texttt{step}, to
advance the distributed application to the next global breakpoint
indicated in the \TeSS file.

On \Deipa console, the distributed application (only \echoclient for
now) starts running and stops before line~17, the \pvmspawn function
call.  When we step into the next global breakpoint, the \echoclient
process is spawned.  As the \texttt{step} command lists all the known
processes before doing any action, its possible to verify that at the
first \texttt{step} there is just one process (corresponding to the
\echoclient program), while at the second \texttt{step} there are
already two processes, meaning the \echoserver was already launched
and is now under control of \Fiddle (and \Deipa).

Since \Ltwo level allows several concurrent clients to connect to
\Fiddle, we also launched a \Fiddle console. Once again, in the
initial state (before the \pvmspawn), there is only one process (with
TID~1), but in the second state there are two processes (with TIDs~1
and~2), corresponding to \echoclient and \echoserver, respectively.

Similarly, in the top left window (the \texttt{\PVM} console), one can
notice that in the initial state there is a single \PVM task, and
after the second step there are two \PVM tasks.

From the observation of the target application with these three
different tools, we may confirm that the new process was launched
correctly and integrated into the debugging environment of \Fiddle and
\Deipa.

\subsubsection{Driving the application behavior}

As stated before, the server program will exit if it receives an even
number, leaving the client blocked forever waiting for a non-existing
reply.  We can change this behavior and force the server to always
reply to the client, no mater it receives an odd or and even, by
forcing the contents of the \emph{value} variable at line~17 of
\echoserver to always be an odd number.

\begin{figure}[htbp]
  \centering
  \includegraphics[width=1.0\textwidth]{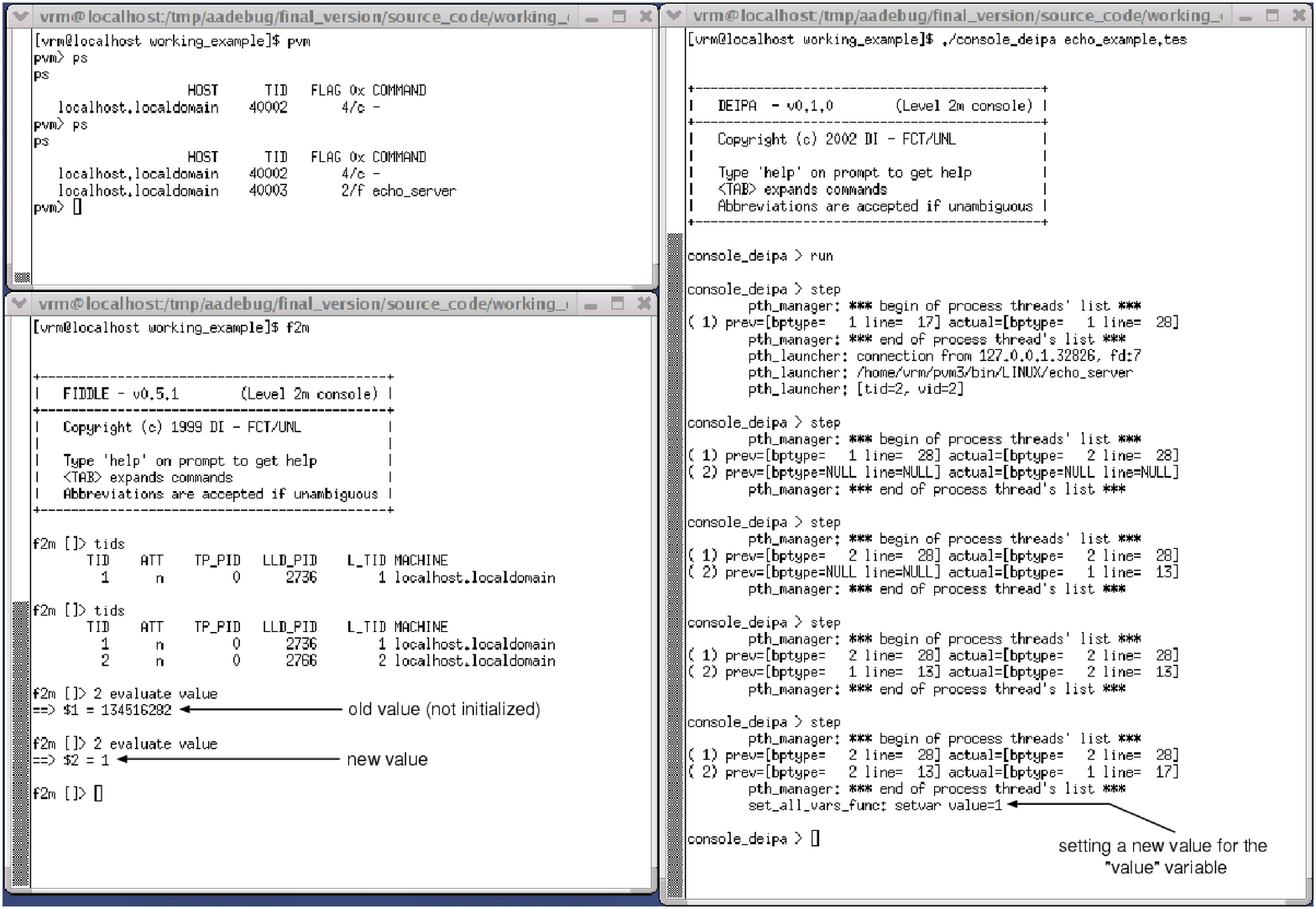}

  \caption{Driving the application behavior}
  \label{fig:forcing}
\end{figure}

The \fref{fig:forcing} shows the same three windows as before.  The
\PVM console om the to left, the \Fiddle \Ltwo console just below, and
\Deipa on the right.

\Deipa was used to control the application until global breakpoint at
line~15 is reached.  At this point, we may use \Fiddle console to
inspect the \emph{value} variable, which is reported to have the
garbage value of 134516282.

Lets step to the next (line~16 of \TeSS file) global breakpoint using
\Deipa, and re-inspect the \emph{value} variable.  Although this
variable has been initialized to zero at line~15 of \echoserver, it
should contain now the value 1.  This is achieved by the special
global breakpoint of \TeSS line~16, which also defines the new
contents of the \emph{value} variable.  The contents of the
\emph{value} variable are checked again on the \Fiddle console, and
the value 1 confirmed.

\subsubsection{Receiving the answer at the client side}

In order to verify that the client always receives a reply message, we
need to execute line~31 of the client's code.  If the server has
exited, the client will remain blocked forever at this line.

\begin{figure}[htbp]
  \centering
  \includegraphics[width=1.0\textwidth]{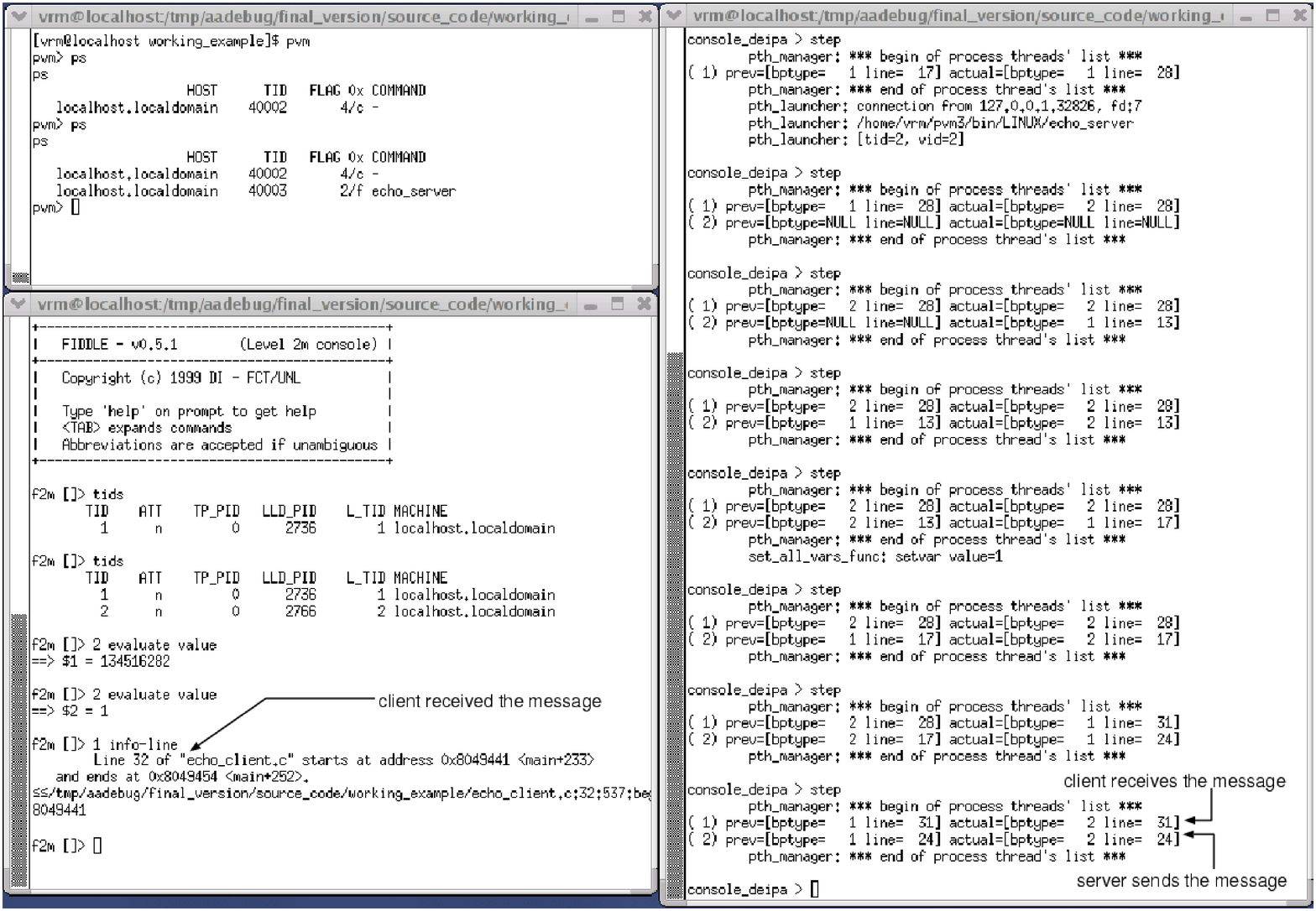}

  \caption{Receiving the answer at the client side}
  \label{fig:receive}
\end{figure}

As shown in \fref{fig:receive}, the \Fiddle console on the left window
informs that the client is stopped at line~32, and \Deipa on the right
window shows that the client stopped after executing line~31 (it is
type~2 global breakpoint).

The above example shows how a set of separately developed tools:
\Deipa and \Fiddle (and also the \PVM console) can be used to drive a
distributed application according to a specification of its desired
behavior, which may differ from the application behavior in an
uncontrolled environment.

\section{Deipa as a Fiddle Client}
\label{sec:deipa}

Above \Ltwo, \Fiddle's API supplies functions to manage multiple
clients on a single \Fiddle session, interacting concurrently with
\Fiddle (and with the target application).  \Deipa, \Fiddle's \Ltwo
console and \FGI are examples of such clients.

\subsection{The DEIPA tool}
\label{deipa_tool}

The \Deipa tool is composed of two programs:
\begin{inparaenum}[i)]
\item \emph{the main program}, which interacts with the user and
  controls the application behavior; and
\item \emph{the launcher}, which captures newly created application
  processes (currently supporting only \PVM) and makes their
  registration on \Deipa and \Fiddle.
\end{inparaenum}

\paragraph{The main program}

From \Fiddle perspective, \Deipa is just another client tool accessing
the target application.  To become a \Fiddle client, and able to
access its services, \Deipa must be linked to the \Fiddle \Ltwo
library and call the registration function available in the library
API.

\paragraph{The launcher}

To capture new \PVM processes and integrate them in the debugging
environment (\Deipa and \Fiddle), the \texttt{PvmTaskDebug} flag must
be activated when the target application spawns a new process, and the
\texttt{PVM\_DEBUGGER} environment variable must specify the
\emph{launcher} location.  In this way, instead of spawning the new
process directly, \PVM spawns the \emph{launcher}, which receives the
name of the process to launch as a command line argument (for more
details on this topic, consult the \PVM documentation~\cite{pvm}).

Once the \emph{launcher} is running, it executes the following steps:
\begin{enumerate}
\item Creates a bi-directional communication channel, for the future
  interactions with \Fiddle;
\item Announces itself to \Fiddle as a node debugger, indicating that
  future interactions should use the communication channel;
\item Announces itself to \Deipa, informing on the new target process
  name and \Fiddle identification number;
\item Redirects its standard I/O channels to the above mentioned
  communication channel;
\item Core image auto-replacement (\texttt{exec}) with a node debugger,
  running the new target process.  The redirected I/O channels are
  inherited on \texttt{exec}'ing, and the node debugger will have its
  I/O redirected to the communication channel.
\end{enumerate}

The \Fiddle API includes a function that allows to
register/incorporate an existing node debugger into the \Fiddle
environment.  The \Deipa \emph{launcher} fits into this category, but
before becoming a node debugger, it supplies some relevant information
to \Deipa.

\subsection{Inside Deipa}
\label{inside_deipa}

\Deipa is structured in two main parts:
\begin{inparaenum}[i)]
\item \emph{the front-end}, currently only a text-oriented user
  interface (\Deipa console) is implemented; and
\item \emph{the back-end}, where all the relevant \Deipa
  functionalities are implemented.
\end{inparaenum}

\paragraph{The Deipa console}

The commands available to the user can be divided in two sets: 
\begin{inparaenum}[i)]
\item the ones whose scope is limited to \Deipa itself, \eg,
  \texttt{open} (load a new \TeSS file), and 
\item those that act upon the target processes, \eg, \texttt{run} and
  \texttt{step} (drive the target application until the first/next
  global breakpoint).
\end{inparaenum}

\paragraph{The Deipa library}

\begin{figure}[htbp]
  \centering
  \includegraphics[width=1.0\textwidth]{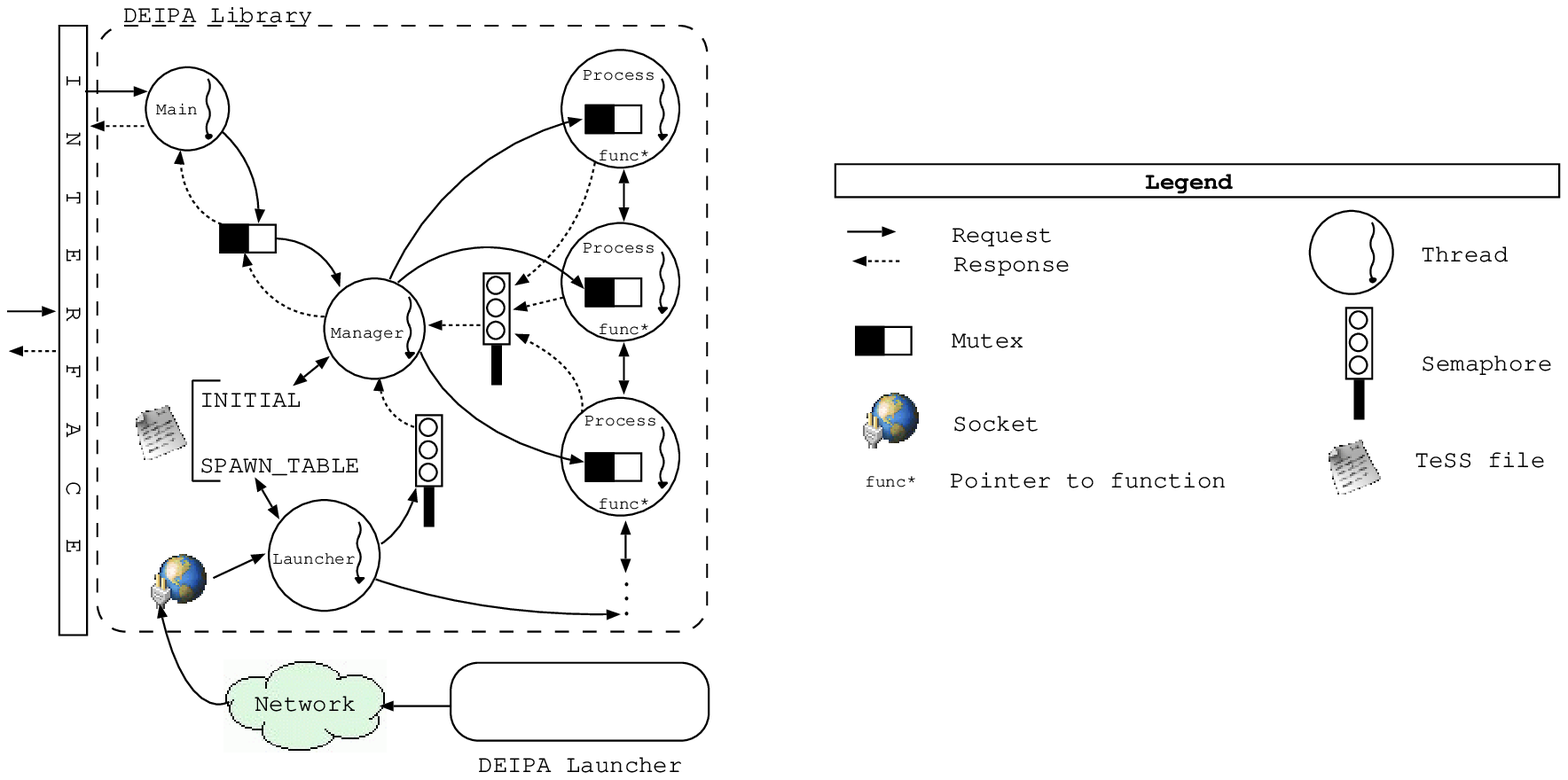}

  \caption{\Deipa implementation}
  \label{fig:deipa-implementation}
\end{figure}

\Deipa library, whose architecture is described in
\fref{fig:deipa-implementation}, uses multiple threads, one for each
target process (the \emph{process threads}) and three internal
management threads.

The process threads are responsible for the control and inspection of
target processes.

The three management threads are:
\begin{inparaenum}[i)]
\item the \mainthread, responsible for receiving requests, verifying if
  a request is valid in the current state of the library, and
  sending the replies;
\item the \managerthread, responsible for managing all the process
  threads; and
\item the \launcherthread, responsible for receiving the registration
  of each new process created by the \pvmspawn and signaled by the
  launcher program.
\end{inparaenum}

Details on \Deipa internals can be found in~\cite{deipa:vrm}.

\section{Conclusions and Future Work}
\label{sec:conclusions}

The \Fiddle engine meets several requirements for the debugging of
multi-thread and multi-process distributed applications. Its
suitability has been evaluated through the development of prototypes
to test the integration of several types of support tools. This
includes the support for graphic debugging interfaces, the integration
of testing and debugging, and the integration of visualization and
debugging tools.

In this paper, we have described how \Fiddle eases the development of
integrated testing and debugging support. This is achieved through a
separate tool, \Deipa, which acts as an intermediary between an
independent testing tool (\STEPS) and the \Fiddle debugging engine.
Furthermore, \Fiddle provides support for concurrent clients, allowing
the user to perform debugging and control of a distributed application
at two distinct levels: the \Deipa level, where global program states
are followed, and at the \Fiddle level, where local process states can
be inspected and modified.

Ongoing work on \Fiddle client tools, such as \FGI and \PADI, will
lead to new experiments on tool integration and cooperation, with the
increased user friendliness given by the graphical user interfaces.
There are plans to do a similar work on \Deipa, replacing its text
console with a much friendlier graphical interface.

Communication between the client tools and \Fiddle, and internally
between \Fiddle components, has been reformulated and changed from a
proprietary data format to \XML.  This usage of the \XML standard will
improve \Fiddle ability to integrate with third party software
packages.

\bibliography{refs}

\end{document}